%% file: main.tex
\documentclass[sigconf]{acmart}

\setcopyright{none}              
\renewcommand\footnotetextcopyrightpermission[1]{} 
\usepackage{algorithmic}
\usepackage{graphicx}
\usepackage{textcomp}
\usepackage{xcolor}
\usepackage{tabularx}
\usepackage{multirow}
\usepackage{float}
\usepackage{multicol}
\usepackage{booktabs}
\usepackage{makecell}
\usepackage{balance}

\begin{document}
\acmConference[]{}{}{}
\title{FormalRTL: Verified RTL Synthesis at Scale}


\author{Kezhi Li}
\affiliation{
    \institution{The Chinese University of Hong Kong}
    \city{Hong Kong}
    \country{China}
}

\author{Min Li}
\affiliation{
    \institution{South East University}
    \city{Nanjing}
    \country{China}
}

\author{Xiangyu Wen}
\affiliation{
    \institution{The Chinese University of Hong Kong}
    \city{Hong Kong}
    \country{China}
}

\author{Shibo Zhao}
\affiliation{
    \institution{South East University}
    \city{Nanjing}
    \country{China}
}

\author{Jieying Wu}
\affiliation{
    \institution{South East University}
    \city{Nanjing}
    \country{China}
}

\author{Junhua Huang}
\affiliation{
    \institution{Huawei Technologies Co. Ltd.}
    \city{Shenzhen}
    \country{China}
}

\author{Qiang Xu}
\affiliation{
    \institution{The Chinese University of Hong Kong}
    \city{Hong Kong}
    \country{China}
}

\begin{abstract}

\input{tex/abstract}
\end{abstract}

\maketitle

\input{tex/introduction}

\input{tex/background}
\input{tex/methodology}
\input{tex/experiment}

\input{tex/conclusion}

{
\clearpage
\bibliographystyle{ACM-Reference-Format}
\bibliography{./references}
}

\end{document}

%% file: tex/abstract.tex
Large language models (LLMs) have demonstrated significant potential in automating hardware synthesis, yet substantial barriers remain for industrial-scale, datapath-centric designs due to ambiguous specifications and a lack of formal correctness guarantees. In this work, we present FormalRTL, a novel end-to-end multi-agent framework that systematically integrates software reference models as formal, executable specifications to guide register-transfer level (RTL) code generation and verification. By tightly coupling planning, synthesis, and formal equivalence checking, FormalRTL achieves scalable and reliable hardware code generation that addresses the critical challenges faced in industrial contexts. The comprehensive evaluation of a new suite of complex industrial-grade benchmarks demonstrates the effectiveness and robustness of our approach. We will open-source the FormalRTL framework and the benchmark suite to facilitate future research in this area.


%% file: tex/introduction.tex
\section{Introduction}
\label{introduction}
As AI workloads continue to scale, the design of computation-intensive chips such as GPUs, NPUs and TPUs has become a major engineering bottleneck~\cite{luo2016dadiannao, choquette2021nvidia, jouppi2023tpu}. This challenge originates from their intricate computational logic 
and relentless pursuit of extreme power, performance and area (PPA) targets~\cite{markidis2018nvidia, arunkumar2017mcm}, which have significantly prolonged the design cycle and increased the development cost of new chips. Nevertheless, with the advancement of AI, large language models (LLMs) have demonstrated impressive capabilities in natural language understanding and code generation~\cite{mastropaolo2023robustness, hui2024qwen2}, which presents new opportunities for automated and agile hardware synthesis.

Recently, a growing body of studies have investigated the use of LLMs to generate hardware register-transfer level (RTL) code~\cite{allam2025asic, liu2024rtlcoder, zhao2025mage, thakur2024verigen}. Approaches based on fine-tuning strategies~\cite{gao2024autovcoder, pei2024betterv, cui2024origen, zhao2025codev} and multi-agent collaboration~\cite{ho2025verilogcoder, yu2025spec2rtl, li2025autosilicon, liu2024rtlcoder} have demonstrated encouraging results on simple open source benchmarks~\cite{thakur2023benchmarking, liu2023verilogeval, liu2024openllm} (e.g. adders and FIFOs with fewer than 100 lines of code). However, these advances highlight the substantial gap between current research prototypes and industrial-scale applications.

We argue that existing methods for industrial-grade RTL synthesis face two fundamental challenges:

\begin{enumerate}

\item \textbf{Large-Scale Design Complexity.} Developing an industrial hardware intellectual property (IP) core typically requires thousands of lines of RTL code. Although planning-based approaches~\cite{liu2024rtlcoder, li2025autosilicon} attempt modular decomposition, their effectiveness is limited by incomplete or ambiguous natural language specifications~\cite{li2025specllm} and the inherent difficulty of LLMs in making sound architectural decisions for complex design.

\item \textbf{Intricate Arithmetic and Logical Structures.} Unlike simple open source benchmarks, industrial IPs often feature deep nested datapaths~\cite{venkataramani2021rapid} and custom data types (e.g. \textit{bfloat16}~\cite{burgess2019bfloat16}, \textit{hifloat8}~\cite{luo2024ascend}). Simulation-based verification~\cite{yan2025assertllm} frequently fails to cover corner cases, particularly in datapath-heavy designs~\cite{blum2002reflections}. 

\end{enumerate}

These limitations in automated hardware generation stem from the inherent difficulty of directly synthesizing RTL code from natural language specifications. Specifications with only natural language often fail to  capture the complex structural, temporal, and arithmetic requirements necessary for hardware implementation~\cite{wei2025vflow}. In conventional industrial practice, hardware design and verification engineers instead rely heavily on software reference models, typically written in C/C++, as executable golden specifications and formal verification artifacts~\cite{clarke2003behavioral, mukherjee2017formal}. Commercial hardware formal verification tools such as Synopsys HECTOR~\cite{synopsys_vcformal_dpv} and Cadence Jasper C Apps~\cite{cadence_jasperc_formal} leverage them to perform high-level equivalence checking (EC), mathematically proving that the final RTL implementation is functionally equivalent to the trusted C/C++ reference model. Despite their industry-standard role, prior LLM-driven hardware synthesis approaches have largely overlooked this valuable resource.

\begin{figure*}[th]
    \centering
    \includegraphics[width=0.85\linewidth]{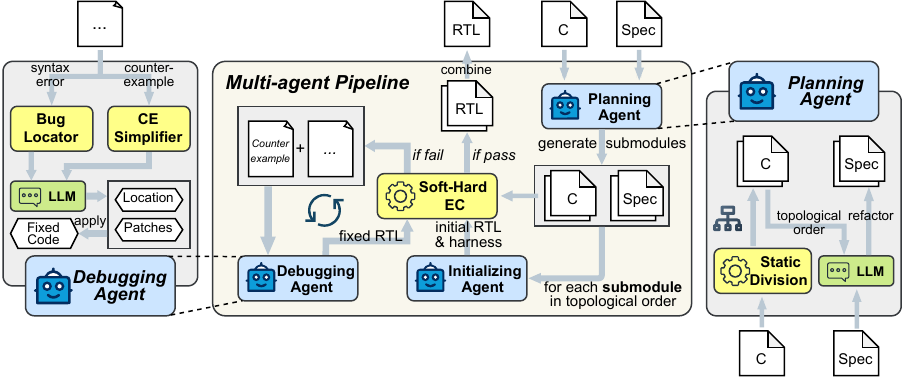}
    \caption{\textbf{The general workflow of FormalRTL.} The \textbf{central pipeline} (middle) takes a C model and specification, then iteratively generates, verifies, and debugs RTL code on a per-submodule basis. The \textit{planning agent} (right) uses C static analysis (AST) to intelligently partition the main task. The \textit{debugging agent} (left) leverages feedback from equivalence checking, using counterexamples to automatically guide an LLM in generating code patches.}
    \label{fig:pipeline}
\end{figure*}

Building on these observations, we introduce FormalRTL (see Fig.~\ref{fig:pipeline}), a novel multi-agent pipeline that systematically addresses the key challenges of industrial-grade hardware synthesis with LLMs. Unlike previous approaches that relied solely on potentially ambiguous natural language specifications, FormalRTL leverages the software reference model as an executable formal specification to guide and anchor the entire synthesis process. 

Specifically, our pipeline begins with a \textbf{planning agent}, which performs static analysis on the C reference model to reliably decompose the complex design into modular subtasks, thereby mitigating architectural ambiguities often found in purely LLM-based planning. For each subtask, the \textbf{initializing agent} generates the RTL code and a verification harness directly anchored to the reference model. To guarantee correctness, FormalRTL applies rigorous EC (i.e. hw-cbmc~\cite{ mukherjee2017formal}) between the software and hardware models. Crucially, the \textbf{debugging agent} closes the loop utilizing counterexamples from the EC tool to efficiently guide automated code refinement and error resolution. This tightly integrated workflow enables scalable RTL generation with formal correctness guarantees, substantially narrowing the gap between LLM-based prototyping and real-world hardware development.

In summary, our main contributions are as follows.
\begin{itemize}
  \item We pioneer a \textbf{software reference model-driven} methodology for agile hardware development, ensuring scalability and correctness by planning via static analysis and equivalence checking. 
  \item We develop an \textbf{end-to-end, multi-agent engine} that automates the entire workflow, from planning, synthesis, and debugging to formal verification.
  \item We construct and release a set of \textbf{industrial-grade benchmarks}, complete with specifications and software reference models. Extensive experiments on these benchmarks show the effectiveness and robustness of our approach.
\end{itemize}

%% file: tex/background.tex
\section{Related Work}
\label{background}

\subsection{LLM for RTL Synthesis}

Early work shows that supervised fine-tuning (SFT) can substantially improve the RTL code generation capabilities of LLMs~\cite{gao2024autovcoder, pei2024betterv, liu2023verilogeval, cui2024origen, zhao2025codev}. To this end, significant effort has focused on generating high-quality training data: CodeV~\cite{zhao2025codev} proposes multi-level summarization data synthesis, OriGen~\cite{cui2024origen} adopts a self-reflection framework with compiler feedback, and BetterV~\cite{pei2024betterv} uses domain-specific instruct-tuning with generative discriminators.

To further enhance reasoning and code generation, recent work has introduced reinforcement learning (RL) mechanisms~\cite{zhu2025codevr1, chen2025chipseek}. ChipSeek-R1~\cite{chen2025chipseek}, for example, integrates direct feedback from simulators, compilers, and electronic design automation (EDA) tools during the training process. Similarly, CodeV-R1~\cite{zhu2025codevr1} annotates its training data with pass/fail feedback from an RTL equivalence checking tool.

However, these SFT and RL approaches remain constrained by the quality of dataset and the scale of the LLMs. Moreover, the fundamental method of direct RTL synthesis from natural language specifications is often unstable and fails to scale to industrial-grade cases. In addition, none of these methods can formally guarantee the correctness of the generated code, even those trained with equivalence checking feedback.

\subsection{Multi-agent Approaches for RTL Synthesis}
Multi-agent collaborative systems offer more scalable and robust RTL synthesis compared to single-LLM methods~\cite{zhao2025mage, li2025autosilicon, ho2025verilogcoder,yu2025spec2rtl, liu2024rtlcoder}. These frameworks typically adopt a three-agent architecture for the core pipeline: task planning, execution/generation, and debugging/optimization. For example, VerilogCoder~\cite{ho2025verilogcoder} employs a graph-based task planning approach using Task and Circuit Relation Graphs (TCRG), while MAGE~\cite{zhao2025mage} introduces a Verilog-state checkpoint mechanism to improve its debugging agent.

However, these frameworks face two fundamental limitations that hinder their industrial applicability. First, their planning relies on unstructured natural language specifications, which are often ambiguous and lack the precision to define complex datapath logic. Second, their debugging depends on simulation-based feedback. This approach is inherently constrained by testbench coverage and cannot provide the formal guarantee of functional correctness—an important requirement for industrial hardware development.

\subsection{Software in Hardware Development}
To overcome the limitations mentioned above, we are motivated by the reliance of the software reference model in the hardware industry, a practice now embedded in many established workflows.

High-level synthesis (HLS) requires designers to write C/C++ programs augmented with pragmas that guide the synthesis process, such as  loop unrolling, pipelining, and interface specification~\cite{o2014xilinx}. This workflow is primarily targeted at FPGA development, enabling rapid prototyping and design-space exploration. Recent advances, including C2HLSC~\cite{collini2024c2hlsc} and HLSPilot~\cite{xiong2024hlspilot}, extend HLS workflows by leveraging LLMs to transform unconstrained C programs into synthesis-friendly C code. 

In contrast, ASIC designs, especially those with intensive algorithm and datapath, rely on software models not only as synthesis inputs but also as golden specifications for architectural exploration and RTL verification. Architects typically encode the intended functionality in C++ or SystemC, which verification engineers then use to generate expected outputs for regressions. ARM, for example, provides cycle-accurate SystemC models of its processor IP~\cite{arm_cycle_model_guide}, while tools such as Cadence Jasper C2RTL~\cite{cadence_jasperc_formal} support formal equivalence between RTL datapaths and high-level C/C++ logic. By grounding verification in these trusted reference models, designers can clarify intent early, expose subtle corner-case bugs, and streamline functional sign-off.



%% file: tex/methodology.tex
\section{Proposed Methods}
\label{sec:methodology}

The general workflow of FormalRTL is illustrated in Fig.~\ref{fig:pipeline}. FormalRTL is designed as a unified multi-agent framework that seamlessly transforms high-level design intent into verified RTL implementations. By explicitly taking both the C reference model and the accompanying specification as inputs, the system orchestrates planning, synthesis, debugging, and formal verification in an automated pipeline. This integrated approach can enable scalable, reliable, and efficient hardware development, ensuring that the generated RTL faithfully adheres to both the intended functionality and the formal specification. In addition, we build a new benchmark that provides both a specification and a software reference model for each case to test our engine.

\subsection{Planning Agent}
The planning agent first partitions the C reference code into subfunctions using static analysis from a C compiler. The compiler analyzes function dependencies via the abstract syntax tree (AST), bundling each C function with its required dependencies (e.g. macros, enum types, and constant variables). Following partitioning, the LLM refactors the main design specification into a targeted specification for each individual submodule based on the C subfunction. 

\begin{figure}[t!]
    \centering
    \includegraphics[width=0.85\linewidth]{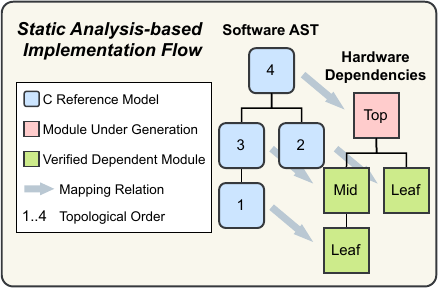}
    \caption{Submodule implementation flow guided by C function dependencies.}
    \label{fig:map}
\end{figure}

The generation process then proceeds strictly following the topological order of the function dependencies derived from the AST. As illustrated in Fig.~\ref{fig:map}, when implementing a given submodule, the agent receives its C reference and all previously verified RTL modules on which it depends.

This partitioning strategy offers three key benefits. First, by explicitly leveraging the functional hierarchy of the C reference model, our planning is more reliable than planning based on the specification alone. Second, the decomposition limits the scope of EC to an acceptable scale, ensuring that the verification tool can efficiently process the task and provide rapid feedback. Third, it enables a bottom-up verification philosophy: each RTL submodule is formally checked for equivalence against its corresponding C function. This incremental verification ensures the correctness of the dependent components, which in turn reduces the debugging burden when building the top modules.

\subsection{Initializing Agent}
The initializing agent takes each submodule's reference model and specification as input and generates both an initial RTL file and a corresponding \textbf{verification harness}. The harness establishes the verification environment for the RTL and C codes by assuming identical inputs and asserting output equivalence~\cite{mukherjee2017formal}. To ensure reliability, we adopt a few-shot prompting strategy that guides the agent in producing the necessary components for a structurally correct harness.

Subsequently, the agent examines whether the submodule requires sequential logic as specified. If so, it determines the appropriate timeframe of the harness, i.e., the exact number of clock cycles required for the hardware to yield final results. This parameter plays a dual role: it governs the activation of transactional equivalence checking~\cite{synopsys_vcformal_dpv, cadence_jasperc_formal, clarke2003behavioral}, while simultaneously serving as a design knob to regulate the depth and granularity of the RTL pipelining. By explicitly encoding the timeframe, pipeline stages are aligned with latency requirements, enabling systematic verification of sequential behavior and offering designers precise control over the generated timing behavior.

\subsection{Debugging Agent}
\begin{figure}[!t]
    \centering
    \includegraphics[width=0.85\linewidth]{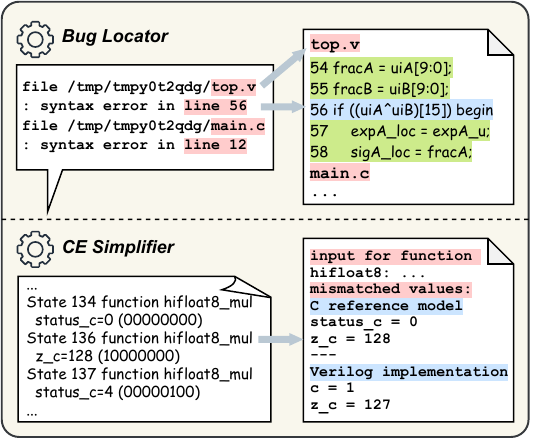}
    \caption{Qualitatively demonstrate bug localization and counterexample simplification.}
    \label{fig:tool}
\end{figure}

The EC tool provides two types of feedback: syntax errors and counterexamples. For syntax errors, we use a \textbf{bug locator} to first identify the error's position. We then extract the code lines surrounding the error to explicitly show the agent the erroneous code. This reduces the LLM's burden of locating the error solely from a line number, which is often difficult without the full code structure. For counterexamples, we use a \textbf{simplifier} that simplifies the report to show only the information about the problematic function and the mismatching signals. This helps the agent quickly identify the source of the logic mismatch. A qualitative demonstration of these two tools is shown in Fig.~\ref{fig:tool}. Finally, the LLM generates the location of the erroneous code and a corresponding patch, which is then applied to the original file to produce the fixed code. 

In this agent, the EC tool plays an important role as both a syntax checker and a counterexample provider. Counterexample-guided debugging is significantly more efficient than binary ``pass/fail'' feedback. Furthermore, the tool guarantees a counterexample upon any equivalence failure. This automated approach effectively replaces the laborious process of manual testbench creation, thereby avoiding the common pitfalls of insufficient coverage and high engineering effort.

\subsection{Benchmark Curation}

Current open source benchmarks for RTL generation often lack industrial-grade specifications and corresponding reference C models. Therefore, a contribution of this work is the collection and curation of a benchmark suite derived from both academic and industrial scenarios. For this initial work, we primarily focus on datapath-intensive modules, deferring the integration of more complex control-logic components to future research.

\textbf{Academic Cases.} Our academic cases are 16-bit floating-point (FP16) operators (e.g. FP16 adder and FP16 mulitiplier) derived from the Berkeley ``softfloat'' repository~\cite{hauser_softfloat}, which provides C implementations of IEEE 754 computation units~\cite{kahan1996ieee}. To prepare inputs for FormalRTL, we first use the static analysis of a C compiler, i.e., Clang,  to merge all scattered subfunctions for a given top-level module into a single file. Then, the file is manually cleaned to ensure that it could be compiled independently. Finally, we author an industrial-style specification for each module, detailing necessary information such as special case handling and sequential requirements to simulate a real-world development process. Each top-level module and its specification constitute a single test case.

\textbf{Industrial Cases.} For industrial cases, we manually implement C reference models for the multiplication and addition functions of Hifloat8~\cite{luo2024ascend}, an 8-bit floating-point datatype featured in the recent Ascend AI Chip~\cite{liao2021ascend}. We also write a detailed specification for this datatype and define its computational behavior with reference to the IEEE 754 standard~\cite{kahan1996ieee}. These cases serve to demonstrate our pipeline's applicability to the agile development of state-of-the-art chip designs.

%% file: tex/experiment.tex
\section{Experiments}
\label{sec:experiment}
\subsection{Experimental Setup}
We build FormalRTL using Langchain~\cite{topsakal2023creating}. After considering factors such as cost-effectiveness and task complexity, we select different base LLMs for different agents. Candidate LLMs are GPT-4.1 and GPT-5. Their application is agent-specific:

\begin{enumerate}
\item The planning agent utilizes GPT-4.1, as this stage primarily involves refining the initial natural language specifications according to the C subfunctions—a task that requires only moderate reasoning ability.
\item The initializing agent employs GPT-5, which is better suited to generate correct, reasonable, and high-quality RTL code. This step requires advanced reasoning capabilities of the model.
\item The debugging agent initially uses GPT-4.1 for rapid code fixes. If persistent issues remain after five iterations, the process escalates to GPT-5 to take advantage of its stronger reasoning for more complex debugging tasks.
\end{enumerate}
We conducted all experiments using LLMs with default configurations.

Furthermore, we use Clang as the compiler choice for static analysis. For verification of software/hardware equivalence, we use an academic C-RTL EC tool, i.e., hw-cbmc~\cite{mukherjee2017formal}.

\subsection{Metrics}
We define several key metrics to evaluate FormalRTL.

\textbf{Initial Success Rate} (ISR, Eq.~\ref{eq:isr}): This is equivalent to the pass@1 metric in code generation benchmarks, which measures the percentage of runs that pass equivalence checking on the first attempt (i.e. without any debugging iterations).

\begin{equation}
   \text{ISR} = \frac{N_{\text{pass}}}{N_{\text{total}}} \times 100\%, 
   \label{eq:isr}
\end{equation}
where $N_{\text{pass}}$ is the number of runs that pass after the initial generation and $N_{\text{total}}$ is the total number of runs, which is 20 in the following experiment.

\textbf{\# of Iterations:} For the debugging phase, we set a maximum iteration limit ($I_{\text{limit}}$) of 20. If the agent fails to fix the error within 20 iterations, the run is marked as a failure. To penalize failure cases, we treat the iteration count for any failed run as double the limit.

We record the Average Fixing Iterations ($I_{\text{avg}}$, Eq.~\ref{eq:iavg}) and the Standard Deviation ($I_{\text{std}}$, Eq.~\ref{eq:istd}).

\begin{equation}
    I_{\text{avg}} = \frac{1}{N_{\text{total}}} \sum_{j=1}^{N_{\text{total}}} I'_{j},
    \label{eq:iavg}
\end{equation}

\begin{equation}
I_{\text{std}} = \sqrt{ \frac{1}{N_{\text{total}}} \sum_{j=1}^{N_{\text{total}}} (I'_{j} - I_{\text{avg}})^2 },
\label{eq:istd}
\end{equation}
where $I'_{j}$ is the adjusted iteration count for the run $j$ as shown in Eq.~\ref{eq:ij}:
\begin{equation}
    I'_{j} = \begin{cases} I_j & \text{if run } j \text{ passes (} I_j \le I_{\text{limit}} \text{)} \\ 2 \times I_{\text{limit}} & \text{if run } j \text{ fails} \end{cases}.
    \label{eq:ij}
\end{equation}

\textbf{Final Success Rate} (FSR): Finally, we record the FSR, which measures the total percentage of designs that were successfully fixed by our agent within the iteration limit.

\subsection{Performance on Module-level Cases}

\input{tab/total_data_twocol}
Table~\ref{tab:submodules} presents the module-level results. Although FormalRTL processes a complete design, including multiple submodules and a specification, as a single input, we report metrics for each individual submodule for better demonstration. The modules labeled ``top'' represent the top-level module of their respective design. To avoid redundancy, the submodules shared between different designs are tested only once. The last column records the RTL code lines of a submodule and the total lines, including all its dependencies. Based on the results, we draw the following conclusions.

\textbf{Our benchmark cases are challenging, reflecting industrial-level complexity,}  and this is demonstrated in two ways. First, the total length of the RTL code shows that the scale of the task is significant, with the hardest case requiring generating more than 1000 lines of code. Second, the ISR for many modules is low, underscoring that current commercial LLMs struggle to produce a correct module on the first attempt.

\textbf{The debugging agent is effective in fixing errors.} All submodules achieve a high FSR within iteration limits. This highlights the efficiency of our counterexample-guided debugging approach.

\textbf{The planning agent effectively distributes implementation complexity, } and this is demonstrated in two ways. First, the agent balances the implementation load. The partitioned submodules are kept within a reasonable and evenly distributed code length, avoiding any single monolithic bottleneck. Second, the bottom-up verification strategy is highly effective. Counter-intuitively, the average ISR for the top-level modules is higher than that of the leaf or mid-level modules. This is because, when generating the top module, the agent has already implemented and formally verified all its dependencies, reducing the implementation burden.

\textbf{FormalRTL is effective in handling sequential logic.} To demonstrate compatibility with sequential circuits, we introduce specific sequential requirements into the specifications of some designs. Generating sequential logic is inherently more difficult and requires correct identification of sequential requirements and the use of transactional equivalence checking. Despite these challenges, the engine still achieved a high FSR for these sequential circuits.

\subsection{Ablation Study of Planning Methods}
To demonstrate the effectiveness of our reference-code-based planning, we compare it against a baseline that represents the current paradigm of specification-only RTL generation~\cite{zhao2025mage, ho2025verilogcoder, yu2025spec2rtl}. Although these contemporary approaches do not support C reference models as a structured input, we design this baseline to strictly emulate their methodology. We first treated the C reference model as an unstructured text within the specification, rather than using it for static analysis. Second, to ensure a fair comparison and control for backend differences, this baseline uses our formal verification engine (i.e. the same initializing and debugging agents) instead of the testbench-based verification or different EDA tools found in their pipelines. This configuration ensures that any observed performance difference is attributable only to the planning method itself.

We quantify the total implementation effort  ($I_\text{{total}}$, Eq.~\ref{eq:itotal}) for FormalRTL by summing the average fixing iterations of the top module and all its dependent submodules, as follows:

\begin{equation}
    I_{\text{total}}^{\text{module}} = I_{\text{avg}}^{\text{module}} + \sum_{i \in \text{dependencies}} I_{\text{avg}}^{i}.
    \label{eq:itotal}
\end{equation}

The FSR of FormalRTL for a given design is defined as the product of the FSR of its top-level module and the FSRs of all its dependent submodules.

The baseline approach must debug the entire design as a single, monolithic task, making it impossible to formally verify individual sub-components against a C reference. Consequently, the iteration budget for the baseline must be compared with our total summed effort ($I_{\text{total}}$). For a fair comparison, we set the maximum fixing iteration limit for each baseline task to be close to the calculated $I_{\text{total}}$ from Table~\ref{tab:submodules}. Specifically, we choose four designs with varying complexity and set the maximum iteration limits as follows: ``f16\_roundToInt'' (Design 1, limit: 5), ``hifloat8\_mul'' (Design 2, limit: 25), ``hifloat8\_add'' (Design 3, limit 25) and ``f16\_add'' (Design 4, limit: 30).

\begin{figure}
    \centering
    \includegraphics[width=\linewidth]{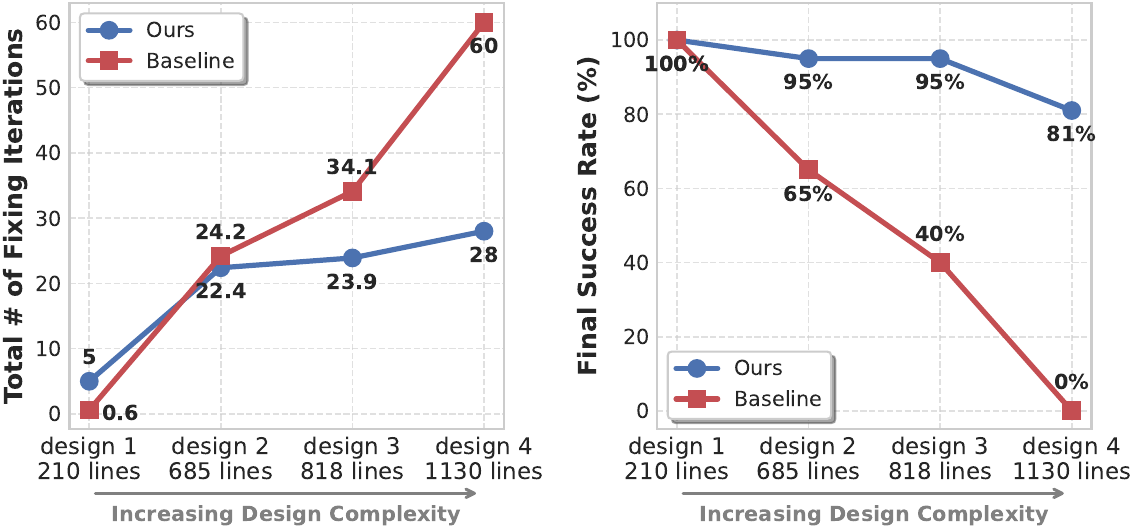}
    \caption{Comparison of planning methods on fixing iterations and FSR.}
    \label{fig:planning}
\end{figure}

As shown in Fig.~\ref{fig:planning}, the performance changes significantly with the design scale. For the small-scale Design 1 (``f16\_roundToInt''), the baseline method requires fewer fixing iterations, as it benefits from generating fixes for all submodules in a single iteration. However, this advantage rapidly diminishes as the complexity increases. For larger designs, the burden of managing the entire design state overwhelms the LLM. In Design 4 (``f16\_add''), the baseline method fails to generate a single correct output with 0\% success rate. In contrast, our reference-code-based planning pipeline maintains a high success rate across all designs, demonstrating superior stability and scalability in handling complex industrial-level RTL generation.

\subsection{Ablation Study of Debugging Tools}
To evaluate the components of the debugging agent, we perform an ablation study on the leaf module ``normalize''. We compare FormalRTL with three baseline configurations: (1) without the bug locator, (2) without the counterexample simplifier, and (3) without counterexample-guided debugging. 

As shown in Table~\ref{tab:debug}, all baseline methods require more fixing iterations and achieve a lower Final Success Rate (FSR) than our complete agent, where the absence of counterexample-guided debugging causes the most significant performance degradation.

\input{tab/ablation_debug}

\subsection{Quality of Result (QoR) Comparison}
We compare the designs generated by FormalRTL with their manual, engineer-optimized counterparts. All designs are evaluated using Yosys~\cite{wolf2013yosys} for synthesis with the Nangate45 library and  OpenROAD~\cite{ajayi2019openroad} for placement and routing.

Table~\ref{tab:ppa} reports two representative cases: ``hifloat8\_mul'' and 
``f16\_mul''. Similar trends are observed in the evaluation of other modules. The LLM-generated RTL typically trails manual designs in area and delay, which is consistent with our emphasis on functional correctness over immediate PPA optimization. Nevertheless, verified RTL provides tangible benefits for engineers: a correct, executable baseline facilitates agile development and offers a reliable foundation for RTL design. Furthermore, a stable golden reference enables systematic PPA iteration and ensures the validity of optimizations.

\input{tab/ppa}

%% file: tab/total_data_twocol.tex

\begin{table*}[htbp]
\caption{Performance of our pipeline on module-level cases.}
\centering
\label{tab:submodules}
\begin{tabular}{@{}cccccccc@{}}
\toprule
\multirow{2}{*}{\textbf{Design}}                                                      & \multirow{2}{*}{\textbf{Module Name}}                                & \multirow{2}{*}{\textbf{Type}} & \multirow{2}{*}{\begin{tabular}[c]{@{}c@{}}\textbf{Initial Success Rate}\\ (pass@1)\end{tabular}} & \multicolumn{2}{c}{\textbf{\# of Iterations}} & \multirow{2}{*}{\textbf{Final Success Rate}} & \multirow{2}{*}{\begin{tabular}[c]{@{}c@{}}\textbf{Module / Total RTL Length}\\ (Average \# of Lines)\end{tabular}} \\ \cmidrule(lr){5-6}
                                                                                      &                                                                      &                                &                                                                                          & $I_{\text{avg}}$      & $I_{\text{std}}$      &                                              &                                                                                                            \\ \midrule
\multirow{13}{*}{\begin{tabular}[c]{@{}c@{}}Softfloat \\ IEEE754\\ FP16\end{tabular}} & countLeadingZeros16                                                  & leaf                           & 90\%                                                                                     & 0.25                  & 0.77                  & 100\%                                        & 48.55 / 48.55                                                                                              \\
                                                                                      & packToF16                                                            & leaf                           & 40\%                                                                                     & 2.50                  & 2.40                  & 100\%                                        & 27.45 / 27.45                                                                                              \\
                                                                                      & shiftRightJam32                                                      & leaf                           & 50\%                                                                                     & 0.90                  & 1.26                  & 100\%                                        & 38.45 / 38.45                                                                                              \\
                                                                                      & propagateNaNF16                                                      & leaf                           & 45\%                                                                                     & 1.75                  & 1.89                  & 100\%                                        & 57.50 / 57.50                                                                                              \\
                                                                                      & normSubnormalF16                                                     & mid                            & 30\%                                                                                     & 5.00                  & 9.05                  & 95\%                                         & 37.70 / 86.25                                                                                              \\
                                                                                      & roundPackToF16                                                       & mid                            & 50\%                                                                                     & 3.45                  & 4.60                  & 100\%                                        & 146.30 / 173.70                                                                                            \\
                                                                                      & addMagsF16                                                           & mid                            & 35\%                                                                                     & 8.85                  & 14.04                 & 85\%                                         & 345.15 / 576.35                                                                                            \\
                                                                                      & subMagsF16                                                           & mid                            & 5\%                                                                                      & 9.25                  & 8.69                  & 95\%                                         & 370.00 / 700.25                                                                                            \\
                                                                                      & f16\_le                                                              & top                            & 50\%                                                                                     & 3.15                  & 8.63                  & 95\%                                         & 62.05 / 62.05                                                                                              \\
                                                                                      & f16\_roundToInt                                                      & top                            & 80\%                                                                                     & 0.70                  & 1.58                  & 100\%                                        & 124.70 / 209.60                                                                                            \\
                                                                                      & f16\_add                                                             & top                            & 65\%                                                                                     & 1.00                  & 1.84                  & 100\%                                        & 46.15 / 1129.95                                                                                            \\
                                                                                      & f16\_sub                                                             & top                            & 90\%                                                                                     & 0.30                  & 1.10                  & 100\%                                        & 43.75 / 1127.60                                                                                            \\
                                                                                      & f16\_mul                                                             & top                            & 15\%                                                                                     & 7.30                  & 8.45                  & 95\%                                         & 209.85 / 565.75                                                                                            \\ \midrule
\multirow{10}{*}{HiFloat8}                                                            & calculate\_dot\_and\_m                                               & leaf                           & 75\%                                                                                     & 0.55                  & 1.12                  & 100\%                                        & 48.25 / 48.25                                                                                              \\
                                                                                      & normalize                                                            & leaf                           & 15\%                                                                                     & 6.40                  & 8.88                  & 95\%                                         & 67.50 / 67.50                                                                                              \\
                                                                                      & lzc                                                                  & leaf                           & 35\%                                                                                     & 1.45                  & 1.63                  & 100\%                                        & 24.10 / 24.10                                                                                              \\
                                                                                      & decode\_hifloat8                                                     & leaf                           & 50\%                                                                                     & 3.30                  & 4.83                  & 100\%                                        & 134.40 / 134.40                                                                                            \\
                                                                                      & encode\_hifloat8                                                     & mid                            & 25\%                                                                                     & 4.20                  & 5.14                  & 100\%                                        & 153.40 / 201.65                                                                                            \\
                                                                                      & round\_half\_to\_away                                                & mid                            & 45\%                                                                                     & 4.75                  & 6.25                  & 100\%                                        & 140.75 / 189.00                                                                                            \\
                                                                                      & hifloat8\_add                                                        & top                            & 70\%                                                                                     & 0.45                  & 0.80                  & 100\%                                        & 257.00 / 818.35                                                                                            \\
                                                                                      & \begin{tabular}[c]{@{}c@{}}hifloat8\_add\\ (sequential)\end{tabular} & top                            & 20\%                                                                                     & 16.00                 & 16.25                 & 70\%                                         & 420.64 / 981.99                                                                                            \\
                                                                                      & hifloat8\_mul                                                        & top                            & 95\%                                                                                     & 0.05                  & 0.22                  & 100\%                                        & 147.05 / 684.35                                                                                            \\
                                                                                      & \begin{tabular}[c]{@{}c@{}}hifloat8\_mul\\ (sequential)\end{tabular} & top                            & 60\%                                                                                     & 9.65                  & 15.67                 & 80\%                                         & 220.10 / 757.35                                                                                            \\ \bottomrule
\end{tabular}
\end{table*}

%% file: tab/ablation_debug.tex
\begin{table}[!t]
\caption{Ablation study of the Debugging Agent.}
\label{tab:debug}
\centering
\begin{tabular}{@{}ccc@{}}
\toprule
\textbf{Methods}        & \textbf{Average \# of Iterations} & \textbf{FSR}  \\ \midrule
w/o bug locator         & 14.15                             & 80\%          \\
w/o CE simplifier       & 9.55                              & 90\%          \\
w/o CE-guided debugging & 21.30                             & 55\%          \\
\textbf{FormalRTL}      & \textbf{6.40}                     & \textbf{95\%} \\ \bottomrule
\end{tabular}
\end{table}

%% file: tab/ppa.tex
\begin{table}[!t]
\centering
\caption{QoR Comparison with Manual Designs.}
\label{tab:ppa}
\begin{tabular}{@{}lcccc@{}}
\toprule
             & \multicolumn{2}{c}{\textbf{hifloat8\_mul}} & \multicolumn{2}{c}{\textbf{f16\_mul}}    \\ \cmidrule(l){2-5} 
             & Area ($\mu m^2$)  & Delay ($ns$)  & Area ($\mu m^2$) & Delay ($ns$) \\ \midrule
FormalRTL & 1015              & 3.29          & 1629             & 3.54         \\
Engineer  & 743               & 1.78          & 1343             & 2.50         \\ \bottomrule
\end{tabular}
\end{table}

%% file: tex/conclusion.tex
\section{Conclusion and Future Work}
\label{conclusion}
This paper presents FormalRTL, a scalable multi-agent engine for LLM-based RTL synthesis. Our pipeline utilizes software reference models to guide the generation of formally verified hardware. The system employs static analysis for task planning and formal EC counterexamples to guide automated debugging. Experiments on a new suite of industrial-grade benchmarks indicate that FormalRTL can generate and verify complex and datapath-heavy designs. Despite these advances, several limitations remain and point toward avenues for future work.

\textbf{Scalability to Industrial-Scale RTL.} Our pipeline has been validated on RTL generation tasks exceeding 1000 lines, significantly narrowing the gap between academic prototypes and industrial needs. Although a gap remains for full industrial-scale designs, our effective decomposition strategy enables the construction of verified large-scale  systems from manageable subtasks.

\textbf{Specialized LLMs for Debugging.} While our pipeline is designed to be model-agnostic, its effectiveness heavily relies on the model's ability to understand debugging feedback (e.g. counterexamples). Future work could focus on training small and specialized models specifically for this counterexample-guided fixing task to reduce the reliance on large-scale, proprietary commercial models.

\textbf{Robust Open Source Equivalence Checking.} We observe a scarcity of actively maintained open source tools for C-RTL EC.  The community would greatly benefit from the development of new open source EC tools that support modern RTL syntax and offer more efficient performance.